\documentclass[dvipsnames,letterpaper, 10 pt, conference]{ieeeconf}

\IEEEoverridecommandlockouts	
\usepackage[table]{xcolor}     

\usepackage{cite}
\usepackage{algorithmic}
\usepackage{textcomp}
\usepackage{graphicx}          
\usepackage{mathrsfs}
\usepackage{tikz}
\usepackage[vlined,ruled]{algorithm2e}
\usepackage{subfigure}
\usepackage{url}
\usepackage{dsfont}
\usepackage{bbm}
\usepackage{booktabs}          
\usepackage{array}
\usepackage{yfonts}
\usepackage[normalem]{ulem}
\usepackage{arydshln,leftidx,mathtools}
\usepackage{multicol}
\usepackage{amsmath}
\usepackage{stfloats}
\usepackage{bbm}
\usepackage{comment}
\usepackage{subfigure}
\usepackage{manyfoot}
\usepackage{listings}
\usepackage{float}
\usepackage[most]{tcolorbox}
\usepackage{placeins}
\usetikzlibrary{positioning,arrows.meta,shapes.misc}
\usepackage{pifont} 

\definecolor{codegreen}{rgb}{0,0.6,0}
\definecolor{codegray}{rgb}{0.5,0.5,0.5}
\definecolor{codepurple}{rgb}{0.58,0,0.82}
\definecolor{backcolour}{rgb}{0.95,0.95,0.92}
\definecolor{ourshade}{gray}{1.92}

\lstdefinestyle{mystyle}{
    backgroundcolor=\color{backcolour},   
    commentstyle=\color{codegreen},
    keywordstyle=\color{magenta},
    numberstyle=\tiny\color{codegray},
    stringstyle=\color{codepurple},
    basicstyle=\ttfamily\footnotesize,
    breakatwhitespace=false,         
    breaklines=true,                 
    captionpos=b,                    
    keepspaces=true,                 
    numbers=left,                    
    numbersep=5pt,                  
    showspaces=false,                
    showstringspaces=false,
    showtabs=false,                  
    tabsize=2
}

\definecolor{myBlue}{RGB}{49,130,189}

\DontPrintSemicolon
\SetKw{KwContinue}{continue}
\SetKw{KwAnd}{and}
\SetKw{KwRet}{return}
\SetKwInput{KwProc}{Procedure}

\usepackage{tabularx}
\usepackage{multirow}
\usepackage{makecell}

\newcommand\aamsout{\bgroup\markoverwith{\textcolor{violet}{\rule[0.5ex]{2pt}{1pt}}}\ULon}

\DeclareFontFamily{U}{stix2bb}{}
\DeclareFontShape{U}{stix2bb}{m}{n} {<-> stix2-mathbb}{}

\DeclareSymbolFont{bbold}{U}{bbold}{m}{n}
\DeclareSymbolFontAlphabet{\mathbbold}{bbold}

\newcommand\oprocendsymbol{\hbox{$\square$}}
\newcommand\oprocend{\relax\ifmmode\else\unskip\hfill\fi\oprocendsymbol}

\tcbset{
  panel/.style n args={2}{ 
    enhanced,
    arc=10pt,                 
    boxrule=0.5pt,            
    colframe=#1!80!black,
    colback=#1!12,            
    colbacktitle=#1!25,       
    coltitle=black,
    fonttitle=\bfseries\normalsize,
,
    title={#2},
    title filled,             
    titlerule=1.2pt,          
    toptitle=2mm, bottomtitle=2mm,
    left=10pt,right=10pt,top=10pt,bottom=10pt,
    before skip=8pt, after skip=10pt,
  }
}

\newtcolorbox{TaskBox}[1]{panel={blue}{#1}}
\newtcolorbox{EnvBox}[1]{panel={green}{#1}}

\newtcolorbox{StageBox}{
  enhanced,
  colback=white,
  colframe=black!60,
  boxrule=0.8pt,
  arc=6pt,
  left=8pt,right=8pt,top=6pt,bottom=6pt,
  before skip=6pt, after skip=8pt,
}

\newcommand*{\QEDA}{\hfill\ensuremath{\blacksquare}}%

\DeclareNewFootnote{R}[roman]

\graphicspath{{figs/}}

\makeatletter
\let\NAT@parse\undefined
\makeatother
\usepackage[colorlinks,urlcolor=blue, linkcolor=blue, citecolor=blue]{hyperref}

\begin{document}

\title{\LARGE \bf VeraGrid-Agent: Tool-Augmented LLMs for Distribution Optimal Power Flow at the Grid Edge}


\author{Shivanshu~Tripathi, Hamed Mohsenian-Rad, and
  Maziar~Raissi \thanks{S.~Tripathi and H.~Mohsenian-Rad are with the Department of
    Electrical and Computer Engineering, and M.~Raissi is with
    the Department of Mathematics at the University of
    California, Riverside,
    \href{mailto:strip008@ucr.edu}{\{\texttt{strip008}}, \href{mailto:hamedrad@ucr.edu}{\texttt{hamedrad}}, \href{mailto:maziarr@ucr.edu}{\texttt{maziarr\}@ucr.edu}}. We acknowledge the use of Google Gemini for generating Fig.~\ref{fig: example_112} and language refinement.}}

\maketitle
\pagestyle{empty}
\thispagestyle{empty}


\begin{abstract}
Language models have demonstrated remarkable success in solving a wide range of tasks. However, answering complex scientific questions about the power flow  often requires solving the distribution optimal power flow (D-OPF) problem. These questions call for numerical solvers and simulators, as linguistic reasoning from parametric knowledge often gives incorrect answers.
In this work, we present VeraGrid-Agent, a tool-augmented LLM that autonomously 
writes the simulator input, executes the open-source VeraGrid
solver, and reads the solver output before answering. To evaluate performance, we introduce VeraGrid-MCQ-150, a set of deterministic, expert template driven, $150$ multiple-choice questions. We evaluate the performance under two regimes: (i) no-tool
reasoning and (ii) agent (LLM with simulator access). Without
tools, every model performs with an accuracy of
$42.7\%$--$49.3\%$. However, with VeraGrid-Agent, accuracy increases to $97.3\%$--$100.0\%$. We also do a failure-mode analysis to show that the few remaining errors arise from 
wrong interpretations during multi-step reasoning, rather than any failure in the simulators execution.
\end{abstract}


\section{Introduction}\label{sec:introduction}

\begin{figure}[!t]
  \centering
  \includegraphics[width=1.04\columnwidth]{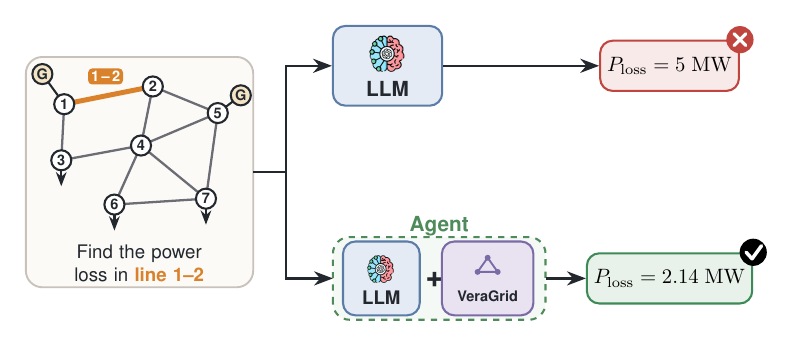}
  \caption{The figure compares a standalone LLM with an agent given the same grid-edge query for a given circuit configuration. The standalone LLM answers from memory and returns an incorrect loss value, whereas the agent invokes the VeraGrid power-flow solver and computes the correct answer.}
  \label{fig: example_112}
\end{figure}

Large language models (LLMs) have advanced rapidly in code generation and multi-step reasoning. These capabilities have encouraged
their adoption in scientific and engineering domains, including power-systems. The electric grid is changing, with distribution feeders now hosting large amounts of distributed energy resources (DERs) \cite{bollen2011dg}. We built an agent that can solve complex questions that are not primarily linguistic, requiring the agents to call the solver tool. The questions include the rise of voltage at the PV node, the capacity of the bus, the curtailment of DERs, the power flow at the substation, or the feeder losses.  
When an LLM is asked such 
questions without any access to tools, it is forced to guess, producing outputs that are numerically incorrect, as illustrated in Fig~\ref{fig: example_112}. 

We consider a DER dispatch problem at the grid-edge and use AC-OPF to solve the problem. For a given network of buses, branches, loads, and generators, AC-OPF
finds the generator dispatch and bus voltages that minimize the total generation cost while
satisfying non-linear AC power-flow equations and the operating limits. This problem is 
non-convex and NP-hard in general~\cite{bienstock2019} and is solved by 
specialized packages such as the open-source VeraGrid simulator. 

This gap has motivated the use of tool-augmented LLMs that connect language
reasoning with calls to external software. Therefore,  rather than answering from
parametric memory, the model writes the input, executes the solver,
reads back the quantities it needs from the solution, and only then commits to an answer. The question we address is whether this pattern is sufficient to measure progress in a way that 
separates genuine computation from memorization and guessing.


\subsection{Related Work}\label{sec:related}

Several works extend the capabilities of LLMs by enabling them with external tools. Toolformer showed that language models can self-supervise to learn when to call tools~\cite{schick2023toolformer}. ReAct showed that connecting LLMs with tool actions improves multi-step decision making and factual grounding~\cite{yao2023react}.  Other agentic frameworks, such as MRKL systems~\cite{karpas2022mrkl}, HuggingGPT~\cite{shen2023hugginggpt} and Gorilla~\cite{patil2023gorilla} showed that LLMs can orchestrate diverse tools, models, and APIs to solve tasks beyond their parametric knowledge.
Our work follows this paradigm by equipping an LLM with a solver to get reliable answers for complex power-flow questions that cannot be recovered from parametric knowledge alone.

The use of LLMs is rapidly expanding in power system applications \cite{llm_power_survey2025}. 
Recent works like ChatGrid~\cite{chatgrid2024} and ElecBench~\cite{elecbench2024} have used retrieval-augmented LLMs to answer questions about power-system. However, these works focus on the textual correctness and usefulness of the generated response, rather than numerical understanding in simulator derived ground truth. More recently, GridMind~\cite{gridmind2025} moved toward executable power-system analysis by coupling LLMs with deterministic solvers. Our work complements these efforts by introducing an MCQ-based evaluation scheme to score different LLMs.  
PowerAgentBench \cite{poweragentbench2025} evaluates tool-using agents in steady-state workflows, such as contingency analysis and mitigation using hidden physical validation. The benchmark in \cite{psagentbench2026} evaluates a broad set of tasks, including power flow, protection, and stability, with deterministic evaluators and held-out test cases~\cite{psagentbench2026}. The work \cite{proopf2026} shows that current models struggle in complicated mathematical analysis. Our work introduces a VeraGrid-MCQ-150, a benchmark that evaluates the performance of models, and their ability to correctly retrieve quantitative information from a solved network through $150$ MCQ questions. 

General scientific and mathematical benchmarks, such as MATH and SciBench~\cite{hendrycks2021math,wang2024scibench,zheng2024mcq} have driven progress in quantitative reasoning but rely on static problem sets that are susceptible to memorization. Our generator addresses
this by making every answer a deterministic function of a
specific solved record, i.e., the same template yields a different ground
truth under different network configurations. Recent
work~\cite{donon2020mlopf} trains a machine learning model to solve OPF, rather than approximating the solver. Instead, VeraGrid-Agent directly calls a deterministic solver, giving exact answers over guessed answers. 

\subsection{Contributions}

This paper makes three main contributions. First, we develop a tool-augmented architecture
    for D-OPF analysis in which the LLM writes the feeder and DER simulator input, executes
    VeraGrid, and reads only the result sections relevant to the 
    question, keeping the context compact and every answer traceable to the solver
    output. Second, we provide a deterministic, template-driven benchmark of $150$ MCQs at easy, medium, and hard difficulty levels, with verified ground truth of the simulator that is reproducible and dependent on the feeders and its DERs. Third, we conduct an empirical study consisting of seven LLMs under the no-tool and agent regimes,
    showing that having access to the tools substantially raises accuracy. We also discuss where the tool helps the most and analyze the failure modes of the agent. 

\section{VeraGrid-Agent}\label{sec:architecture}

\subsection{Task overview}
The D-OPF with DER dispatch for a feeder graph $\mathcal{G}=(\mathcal{N},\mathcal{E})$ is expressed as:
\begin{subequations}\label{eq:dopf}
\begin{align}
\min_{V_i,p_i,q_i}\;
 &\sum_{\mathcal{E}}|I_{ij}|^{2}
  +\lambda_{c}\!\sum_{\mathcal{D}}(\bar{p}_i\!-\!p_i)
  +\lambda_{v}\!\sum_{\mathcal{N}}|V_i|^{2} \label{eq:dobj}\\
\text{s.t.}\;
 &S_i^{\mathrm{G}}\!+\!s_i\!-\!S_i^{\mathrm{L}}=V_iI_i^{*},\;\;
  \mathbf{I}=\mathbf{Y}\mathbf{V} \label{eq:dpf}\\
 &0\le p_i\le\bar{p}_i,\quad |s_i|\le\bar{s}_i \label{eq:der}\\
 &\underline{V}_i\le|V_i|\le\overline{V}_i,\quad
  |S_{ij}|\le\bar{S}_{ij} \label{eq:lim}
\end{align}
\end{subequations}

$\mathcal{N}$, $\mathcal{E}$, $\mathcal{D}\subseteq\mathcal{N}$ are the sets of buses,
branches, and DER buses; $V_i=|V_i|e^{\jmath\theta_i}$, $I_i$, and $I_{ij}$ the bus
voltage, injected current, and branch current; $\mathbf{Y}$ the admittance matrix;
$S_i^{\mathrm{G}}$, $S_i^{\mathrm{L}}$ the substation and load power; $s_i=p_i+\jmath q_i$
the DER injection with limits $\bar{p}_i$ (available power) and $\bar{s}_i$ (inverter
rating); and $\lambda_c,\lambda_v>0$ the objective weights. The
Problem~\eqref{eq:dopf} is non-convex in general~\cite{bienstock2019}, so
practical instances are solved numerically. Let $\Phi$ denote the solution
operator implemented by VeraGrid simulator and let $R=\Phi(\mathcal{G})$ denote the output of the simulator containing records of voltages, flows, dispatches, and cost. Our task is to
answer queries in natural-language about $R$ without revealing $R$ directly to
the model. To solve the problem, we design a VeraGrid-Agent that autonomously writes, executes, and interprets AC-OPF simulations. 

We evaluate the agent by introducing the VeraGrid-MCQ-150 dataset, a rule-based
benchmark of $150$ four-option multiple-choice questions (MCQs) with single correct answer generated autonomously
from generation rules specified by domain experts. 


\subsection{Performance metric}\label{sec:task}

Consider each MCQ with correct label
$\ell^\star$ and the predicted label by 
LLM or the agent ($\mathcal{A}$) as 
$\hat{\ell}$. Missing or malformed responses count as incorrect. Performance
is evaluated as:
\begin{equation}\label{eq:accuracy}
    \mathrm{Acc}(\mathcal{A}) =
    \frac{1}{|\mathcal{Q}|} \sum_{i=1}^{|\mathcal{Q}|}
    \mathbbm{1}\bigl[\hat{\ell}_i = \ell^\star_i\bigr].
\end{equation}
where $|\mathcal{Q}|$ denotes the number of questions. 

\begin{figure}[!t]
  \centering
  \includegraphics[width=1\columnwidth]{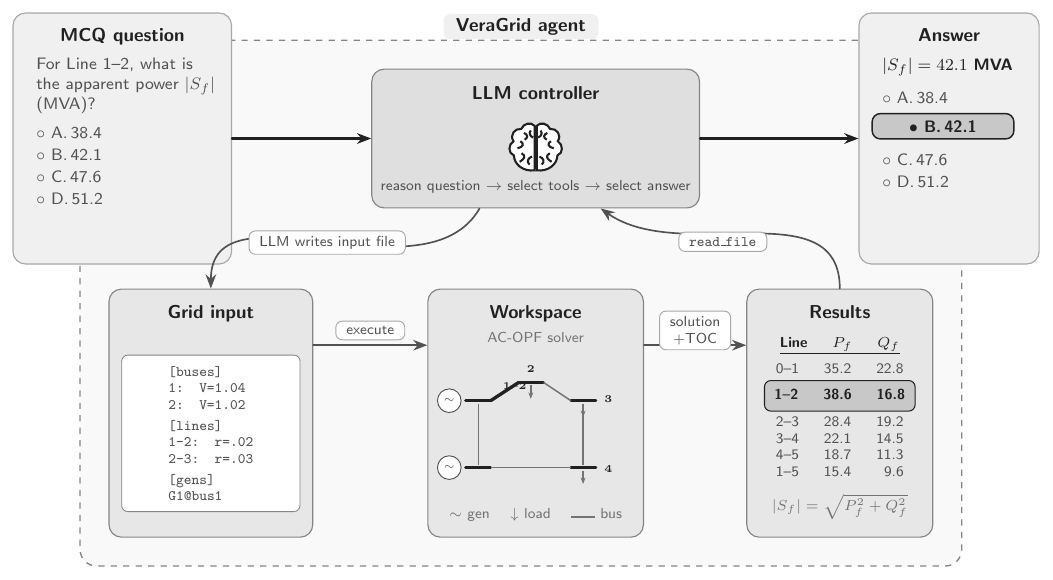}
  \caption{The figure shows the VeraGrid-Agent workflow. For each question, the LLM
  receives the feeder and DER description and MCQ query, then iteratively calls the tools before emitting the final answer.}
  \label{fig: example_1}
\end{figure}

\subsection{Tool Interface and Control Loop}\label{subsec:tools}

Fig.~\ref{fig: example_1} summarizes the workflow of VeraGrid-Agent. Each question is solved
in an isolated workspace. The LLM may call these tools:
\texttt{read\_file} for workspace inspection;
\texttt{write\_file} to author the JSON input describing the feeder and DER configuration; and
\texttt{execute\_veragrid} to run the simulator. A deliberate design 
choice is that the solver tool does \emph{not} return the full solution.
It returns execution metadata and a table of contents for the results 
file. This helps the agent to have targeted reads, keep the context
window compact, and make every reported value traceable to a specific
output line.
The agent follows the ReAct framework~\cite{yao2023react}, at each step,
it either calls a tool or submits a final answer. 
To prevent runaway 
loops in cases where the model keeps issuing tool calls without converging on 
an answer, we cap the number of iterations at $T_{\text{max}}$, the agent returns 
failure if this budget is exhausted. In practice, the great majority of queries are 
resolved in a few iterations.

\subsection{System prompt}
The prompt consists of:  system prompt, 
and query prompt. The system prompt defines the agent's 
role, available tools, the VeraGrid input schema, the feeder and DER details, and answer format. 
It is fixed across interactions and encodes the operational rules of the agent. 
This defines the feeder and DER configuration and changes whenever the configuration changes. The query prompt contains the user query as MCQ.
The system prompt provided to the LLM controller is:
\begin{tcolorbox}[
    breakable,
    enhanced jigsaw,
    colback=gray!7,
    colframe=gray!35,
    boxrule=0.4pt,
    arc=2pt,
    left=6pt,
    right=6pt,
    top=5pt,
    bottom=5pt,
    width=\linewidth
]
\small\ttfamily
You are a VeraGrid agent. \\

=== TOOLS ===\\
- read\_file(path, start, end): read a file \\
- write\_file(name, content): overwrite file\\
- run\_veragrid(input): run VeraGridEngine\\

=== WORKFLOW RULES ===\\
- use tools for every quantitative query if available\\ 
- on failure, inspect error, fix and retry\\

=== INPUT SCHEMA (JSON) ===\\
\{ "name", "baseMVA", "baseKV",\\
\hspace*{1em}"buses": [\{id, type, Vm, Pd, Qd\}],\\
\hspace*{1em}"branches": [\{from, to, r, x, b, rateA\}],\\
\hspace*{1em}"ders": [\{bus, kind, Pavl, Sinv, Qmin, Qmax\}],\\
\hspace*{1em}"substation": \{bus, Vset\} \}\\

=== FEEDER DETAILS ===\\
Substation @ Bus 0: V = 1.00 pu, 12.47 kV\\
Bus 17 (PQ+PV): Pd = 0.10 MW, Qd = 0.06 MVAr, Pavl = 0.60 MW, Sinv = 0.66 MVA\\
Bus 30 (PQ+storage): Pd = 0.20 MW, Sinv = 0.50 MVA\\
Line 0-1: r = 0.05, x = 0.11, rateA = 6.0 MVA\\
Voltage band: 0.95-1.05 pu \\
\ldots\\

=== ANSWER FORMAT ===\\
Respond with ONLY one option (A, B, C, D).
\end{tcolorbox}

\section{Rule based Dataset generation}\label{sec:mcq-generation}

The VeraGrid-MCQ-150 is produced by a deterministic, template-driven generator whose ground truth is computed directly by the VeraGrid simulator. Given the ground truth answers from the simulator $R=\Phi(\mathcal{G})$, each template in the library
specifies (i) a question, (ii) a formula that computes the
correct value from $R$, and (iii)~rules for constructing the three distractors. The benchmark contains three difficulty levels: easy, medium,
and hard, with $50$ questions per livel. Each question is constructed so that it cannot be answered from the LLM’s parametric memory alone. The generated questions are dependent on the feeder and DER configuration, i.e., different configuration yields different sets of questions. 

The three expert-defined difficulty tiers span direct
retrieval, single-step derivation, and multi-step quantitative
reasoning. The benchmark comprises $33$ templates in total: $8$ Easy,
$10$ Medium, and $15$ Hard. 

\begin{itemize}
  \item {Easy}: The questions include direct lookup, counting, and
  identification on the solved feeder. Examples include node voltage magnitudes, DER inventory, feeder element counts, extrema (e.g., the worst voltage node or the most heavily loaded transformer), solver convergence status, and feeder base voltage and base power. In this, we do not require any
  arithmetic beyond retrieval.

  \item {Medium}: These questions require single-step derivations to retrieve answers, such as per-unit conversion, nodal power
  balance at DER node, line impedance magnitude $|Z| = \sqrt{r^2 + x^2}$, DER apparent
  power and power factor, generator utilization and reserve, curtailment faction, the $R/X$
  ratio, the feeder loss ratio, and the voltage rise at DER node.

  \item {Hard}: These questions require multi-step computation,
  cross-table inference, and system-level statistics. These include
  branch loss aggregated from both terminals,
  $P_{\text{loss}} = P_f + P_t$; net reserve power at sub-station, per-unit current magnitude,
  $|I_f| = |S_f| / V_m$; bus voltage-angle differences; DER real/reactive dispatch shares, and
  and thermal margins.
\end{itemize}

The benchmark is designed to evaluate the computation ability of an agent rather than the parametric knowledge. The answer choices are randomly permuted to mitigate the position and selection bias~\cite{zheng2024mcq}. Since every answer is derived from a specific solved network, any change in loads, impedances, or generator limits alters the ground truth. The benchmark is fully reproducible\footnote{Open source dataset and code repository are available: \href{https://github.com/shivanshutripath/Spectrogram-Based-Temporal-Event-Detection}{https://github.com/Shivanshu-11/veragrid-llm-benchmark}}.

\subsection{Evaluation Protocol}\label{subsec:protocol}

We evaluate the framework under two regimes that differ only in tool access, with system prompt and MCQs fixed.
\begin{itemize}
    \item {No-tool}: one chain-of-thought response from the feeder and DER
    description and the question.
    \item {Agent}: the full \textsc{VeraGrid-Agent} loop with
    iteration and self-correction~\cite{yao2023react}.
\end{itemize}

\section{Results}\label{sec:results}


\begin{figure}[!t]
  \centering
  \includegraphics[width=\columnwidth]{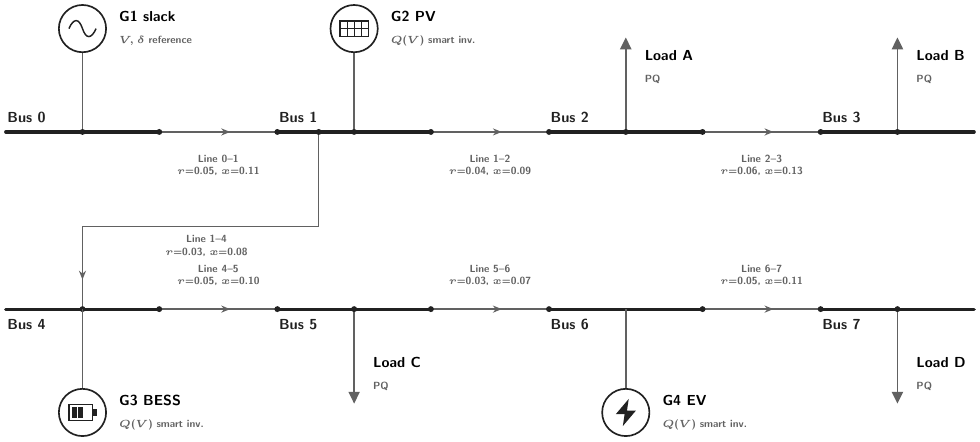}
  \caption{The figure shows a DER-integrated radial distribution feeder with $8$-bus, and $4$-generator used in VeraGrid-MCQ-150. Bus~0 is the slack, buses~1, 4, and~6 host rooftop PV, battery storage, and EV charging with smart inverter reactive control. The remaining buses are PQ loads.}
  \label{fig: case1}
\end{figure}

We consider a DER integrated radial distribution feeder with $8$-buses as shown in Fig.~\ref{fig: case1},
with $185$~MW total load with $100$~MVA base. We consider two regimes: a no-tool baseline in which the model answers from the feeder and DER description alone, and a tool-augmented agent in which the model may invoke the VeraGrid distribution AC-OPF solver and read back the solved state before committing to an answer. 

We compare the performance of several LLM models such as GPT-5.2, Claude~Sonnet~4.5, Gemini~3~Flash,
Grok~4.3, Grok~4.5, Composer~2.5, and Opus~4.8. 

\subsection{Parametric Baseline vs.\ Tool-Augmented Accuracy}

Table~\ref{tab:overall} reports the overall accuracy of the two regimes: without tool access and with VeraGrid-Agent access. 

Without tools, every
model scores only $41.3$--$49.3\%$. However, with the model having access to the VeraGrid-agent, every model scores $97.3$--$100.0\%$.
Gemini~3~Flash, Grok~4.3, and Grok~4.5 reach perfect  $100\%$ accuracy; GPT-5.2 and
Claude~Sonnet~4.5 each miss a single hard question, and Opus~4.8 and
Composer~2.5 misses the correct answer for two and four questions, respectively. 

%

\begin{table}[!t]
\centering
\caption{Accuracy on VeraGrid-MCQ-150 in the no-tool and
tool-augmented agent regimes. The $\Delta$
is the improvement from the no-tool to agent case.}
\label{tab:overall}
\small
\setlength{\tabcolsep}{4pt}
\resizebox{\columnwidth}{!}{%
\begin{tabular}{l l c c c c}
\toprule
Model & Mode & Easy & Medium & Hard & Overall \\
\midrule
\multirow{3}{*}{GPT-5.2}
  & No tool  & $18/50$ & $28/50$ & $18/50$ & $42.7\%$ \\
  & Agent    & $50/50$ & $50/50$ & $49/50$ & $99.3\%$ \\
  & $\Delta$ & $+32$   & $+22$   & $+31$   & $+56.7\%$  \\
\midrule
\multirow{3}{*}{Claude Sonnet 4.5}
  & No tool  & $18/50$ & $34/50$ & $19/50$ & $47.3\%$ \\
  & Agent    & $50/50$ & $50/50$ & $49/50$ & $99.3\%$ \\
  & $\Delta$ & $+32$   & $+16$   & $+30$   & $+52.0\%$  \\
\midrule
\multirow{3}{*}{Gemini 3 Flash}
  & No tool  & $20/50$ & $33/50$ & $21/50$ & $49.3\%$ \\
  & Agent    & $50/50$ & $50/50$ & $50/50$ & $100.0\%$ \\
  & $\Delta$ & $+30$   & $+17$   & $+29$   & $+50.7\%$  \\
\midrule
\multirow{3}{*}{Grok 4.3}
  & No tool  & $24/50$ & $27/50$ & $18/50$ & $46.0\%$ \\
  & Agent    & $50/50$ & $50/50$ & $50/50$ & $100.0\%$ \\
  & $\Delta$ & $+26$   & $+23$   & $+32$   & $+54.0\%$  \\
\midrule
\multirow{3}{*}{Grok 4.5}
  & No tool  & $20/50$ & $28/50$ & $19/50$ & $44.7\%$ \\
  & Agent    & $50/50$ & $50/50$ & $50/50$ & $100.0\%$ \\
  & $\Delta$ & $+30$   & $+22$   & $+31$   & $+55.3\%$  \\
\midrule
\multirow{3}{*}{Composer 2.5}
  & No tool  & $19/50$ & $30/50$ & $18/50$ & $44.7\%$ \\
  & Agent    & $50/50$ & $50/50$ & $46/50$ & $97.3\%$ \\
  & $\Delta$ & $+31$   & $+20$   & $+28$   & $+52.7\%$  \\
\midrule
\multirow{3}{*}{Opus 4.8}
  & No tool  & $19/50$ & $29/50$ & $14/50$ & $41.3\%$ \\
  & Agent    & $50/50$ & $50/50$ & $48/50$ & $98.7\%$ \\
  & $\Delta$ & $+31$   & $+21$   & $+34$   & $+57.3\%$  \\
\bottomrule
\end{tabular}%
}
\end{table}


\subsection{Where Tools Help and Failure mode analysis} 

\begin{figure}[!t]
  \centering
  \includegraphics[width=\columnwidth]{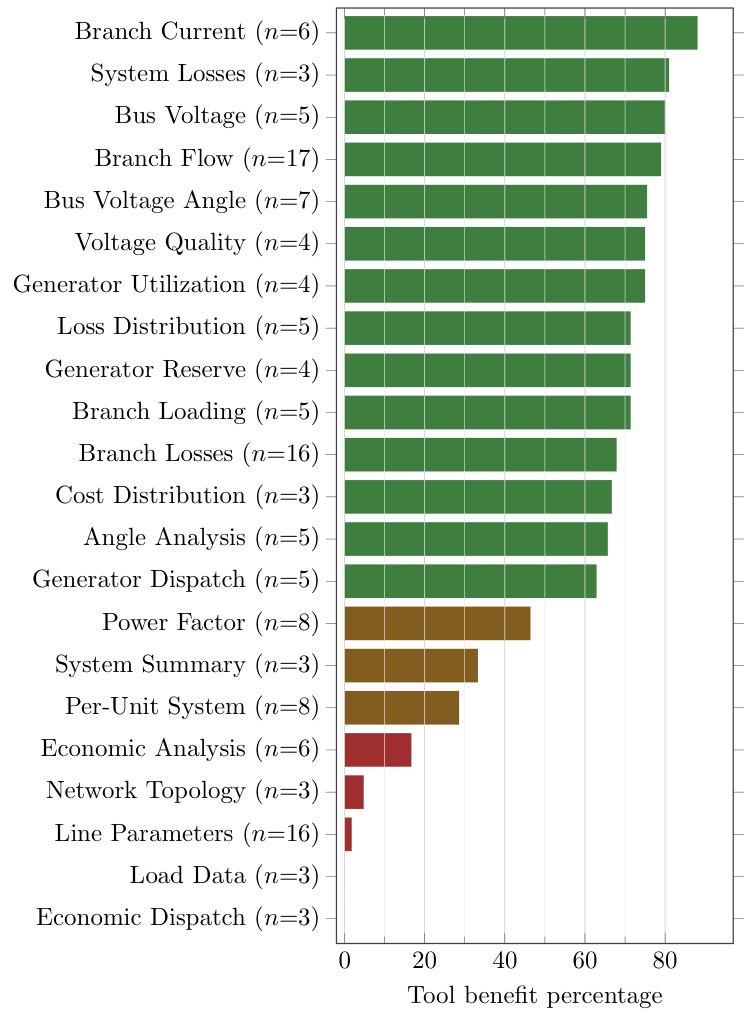}
  \caption{The figure illustrates the improvement from tool augmentation relative to the no-tool baseline, averaged across models on different question category with $n\geq3$. The tool benefit is defined as the difference in accuracy between the agent and no-tool settings, averaged across all models. The green bars indicate substantial gains of $60$--$90\%$, whereas the red bars show only marginal improvements for categories where the no-tool baseline already achieves high accuracy.}
  \label{fig: catbenefit}
\end{figure}

The benefit of the tool is strongly uneven across different categories, as shown in Fig.~\ref{fig: catbenefit}. 
We observe that the questions that are computable from the circuit 
description alone gain little because the no-tool baseline is already high. For example, questions related to line parameters, feeder topology, and
load and DER nameplate data can be answered fairly well without the solver, whereas questions about reserve power flow, feeder losses, and node voltages require it.


Table~\ref{tab:failure} decomposes the error modes for different models. There are no execution errors or timeouts.
Across different
models, the median solved question needs only $1$-$4$ tool calls, which takes an average of 
$17$-$27$~s. Opus~4.8 is the most economical backbone with median call to the $1$ tool in $17$~s, yet not the most accurate, whereas the model with the full accuracy
scorers spends $3$-$4$ calls with $20$-$24$~s. 

\begin{table}[!t]
\centering
\caption{Error decomposition for each model in Agent mode, reported as counts over 150 questions per model. No tool use refers to cases in which the model answered from memory without invoking any external tools. Wrong interpretation denotes cases in which the execution completed successfully, but the model submitted an incorrect answer.}
\label{tab:failure}
\begin{tabular}{l c c c}
\toprule
Model 
      & \makecell{No tool\\use} & \makecell{Wrong\\interpretation} & Total \\
\midrule
GPT-5.2            & $0$ & $1$ & $1$ \\
Claude Sonnet 4.5  & $0$ & $1$ & $1$ \\
Gemini 3 Flash    & $0$ & $0$ & $0$ \\
Grok 4.3          & $0$ & $0$ & $0$ \\
Grok 4.5           & $0$ & $0$ & $0$ \\
Composer 2.5       & $1$ & $3$ & $4$ \\
Opus 4.8           & $0$ & $2$ & $2$ \\
\bottomrule
\end{tabular}%
\end{table}

\subsection{Discussion}

The results provide a comparison of the performances of the agent (LLM+tools) and LLM alone. VeraGrid performs the DER dispatch computations, while the model translates a natural-language query into the appropriate sequence of tool calls and maps the returned quantities to the correct answer. 

For grounded grid-edge question answering, the choice of LLM backbone matters far less than access to computation. In fact, the best models like Opus-4.8 and GPT-5.2 have lower performance than many older models. Once numerical reasoning is provided to the solver, large differences in baseline performance disappear largely. Two observations prove this conclusion. First, the benefits of tool use are highly concentrated. Categories that can already be answered from the prompt, including feeder topology, line parameters, and load data, exhibit little improvement. Second, the failures are primarily semantic rather than the solver's error. Models can execute the solver correctly, yet still produce an incorrect answer by formulating the wrong procedure, such as aggregating branch loss at a single terminal, or confusing DER curtailment with DER dispatch. 


\section{Conclusion}\label{sec:conclusion}

We presented VeraGrid-Agent, a tool augmented LLMs, and the VeraGrid-MCQ-150 benchmark, a reproducible testbed to evaluate the performance of the LLMs in the no-tool and tool augmented settings. Our experiments show that access to external tools improves performance and can solve tasks that cannot be reliably solved using parametric knowledge alone. These results highlight the importance of solver-based agents for numerically intensive engineering. Still several limitations remain as our experiments use only a single radial feeder modeled in balanced form. Future work will extend this framework to larger unbalanced three-phase feeders. Another issue is that the VeraGrid solver takes a lot of time to give the final solution. The future work will reduce this time-taken by the agent to reach the final answer. We will use the reasoning and the response generated by VeraGrid-Agent to fine-tune a small language model to have improved power-system reasoning capabilities without any access to tools.

\bibliographystyle{IEEEtran}
\bibliography{Ref}

@article{bienstock2019,
  author  = {D. Bienstock and A. Verma},
  title   = {Strong {NP}-hardness of {AC} power flows feasibility},
  journal = {Operations Research Letters},
  volume  = {47},
  number  = {6},
  pages   = {494--501},
  year    = {2019}
}

@book{bollen2011dg,
  author    = {Bollen, Math H. J. and Hassan, Fainan},
  title     = {Integration of Distributed Generation in the Power System},
  publisher = {Wiley-IEEE Press},
  year      = {2011}
}

@article{donon2020mlopf,
  author  = {D. Owerko and F. Gama and A. Ribeiro},
  title   = {Optimal Power Flow Using Graph Neural Networks},
  journal = {International Conference on Acoustics, Speech, and Signal Processing (ICASSP)},
  pages   = {5930--5934},
  year    = {2020}
}

@inproceedings{yao2023react,
  author    = {S. Yao and J. Zhao and D. Yu and N. Du and I. Shafran and K. Narasimhan and Y. Cao},
  title     = {{ReAct}: Synergizing reasoning and acting in language models},
  booktitle = {Proc. Int. Conf. Learning Representations (ICLR)},
  year      = {2023}
}

@inproceedings{schick2023toolformer,
  author    = {T. Schick and J. Dwivedi-Yu and R. Dess{\`i} and R. Raileanu and M. Lomeli and E. Hambro and L. Zettlemoyer and N. Cancedda and T. Scialom},
  title     = {Toolformer: Language models can teach themselves to use tools},
  booktitle = {Advances in Neural Information Processing Systems},
  volume    = {36},
  year      = {2023}
}

@article{karpas2022mrkl,
  author  = {E. Karpas and O. Abend and Y. Belinkov and O. Lieber and N. Ratner and Y. Shoham and H. Bata and Y. Levine and K. Leyton-Brown and D. Muhlgay and N. Rozen and E. Schwartz and G. Shachaf and S. Shalev-Shwartz and A. Shashua and M. Tenenholtz},
  title   = {{MRKL} systems: A modular, neuro-symbolic architecture that combines large language models, external knowledge sources and discrete reasoning},
  journal = {arXiv preprint arXiv:2205.00445},
  year    = {2022}
}

@inproceedings{shen2023hugginggpt,
  author    = {Y. Shen and K. Song and X. Tan and D. Li and W. Lu and Y. Zhuang},
  title     = {{HuggingGPT}: Solving {AI} tasks with {ChatGPT} and its friends in {Hugging Face}},
  booktitle = {Advances in Neural Information Processing Systems},
  volume    = {36},
  year      = {2023}
}

@inproceedings{patil2023gorilla,
  author    = {S. G. Patil and T. Zhang and X. Wang and J. E. Gonzalez},
  title     = {Gorilla: Large language model connected with massive {APIs}},
  booktitle = {Advances in Neural Information Processing Systems},
  volume    = {37},
  pages     = {126544--126565},
  year      = {2024}
}

@inproceedings{zheng2024mcq,
  author    = {C. Zheng and H. Zhou and F. Meng and J. Zhou and M. Huang},
  title     = {Large language models are not robust multiple choice selectors},
  booktitle = {Proc. Int. Conf. Learning Representations (ICLR)},
  year      = {2024}
}

@article{hendrycks2021math,
  author  = {D. Hendrycks and C. Burns and S. Basart and A. Zou and M. Zhu and D. Song and J. Steinhardt},
  title   = {Measuring mathematical problem solving with the {MATH} dataset},
  journal = {arXiv preprint arXiv:2103.03874},
  year    = {2021}
}

@inproceedings{wang2024scibench,
  author    = {X. Wang and Z. Hu and P. Lu and Y. Zhu and J. Zhang and S. Subramaniam and A. R. Loomba and S. Zhang and Y. Sun and W. Wang},
  title     = {{SciBench}: Evaluating college-level scientific problem-solving abilities of large language models},
  booktitle = {Proc. Int. Conf. Machine Learning (ICML)},
  volume    = {235},
  pages     = {50622--50649},
  year      = {2024}
}

@article{gridmind2025,
  author  = {H. Jin and K. Kim and J. Kwon},
  title   = {{GridMind}: {LLMs}-powered agents for power system analysis and operations},
  journal = {arXiv preprint arXiv:2509.02494},
  year    = {2025}
}

@article{poweragentbench2025,
  author  = {C. Mylonas and M. Foti and A. Pomarico and M. Duarte and Q. Zhang and E. Varvarigos},
  title   = {{PowerAgentBench-SS}: A benchmark for agentic {AI} in power system steady-state studies},
  journal = {arXiv preprint arXiv:2606.18789},
  year    = {2026}
}

@article{psagentbench2026,
  author  = {S. Trashchenkov},
  title   = {Power systems agent benchmark: Executable evaluation of {AI} agents in electric power engineering},
  journal = {arXiv preprint arXiv:2606.20950},
  year    = {2026}
}

@article{proopf2026,
  author  = {C. Shen and Z. Guo and X. Wan and Z. Yang and Y. Zhang and W. Huang and J. Song and Z. Zhang and M. Sun},
  title   = {{ProOPF}: Benchmarking and improving {LLMs} for professional-grade power systems optimization modeling},
  journal = {arXiv preprint arXiv:2602.03070},
  year    = {2026}
}

@article{elecbench2024,
  author  = {X. Zhou and H. Zhao and Y. Cheng and G. Liang and G. Liu and W. Liu and Y. Xu and J. Zhao},
  title   = {{ElecBench}: A power dispatch evaluation benchmark for large language models},
  journal = {arXiv preprint arXiv:2407.05365},
  year    = {2024}
}

@inproceedings{chatgrid2024,
  author    = {M. Ni and J. Zhang and C. Fu and J. Wang and X. Ning and S. Li},
  title     = {{ChatGrid}: Intelligent knowledge {Q\&A} for power dispatching control based on large language models and retrieval-augmented generation},
  booktitle = {2024 IEEE 7th Information Technology, Networking, Electronic and Automation Control Conference (ITNEC)},
  volume    = {7},
  pages     = {921--925},
  year      = {2024}
}

@article{llm_power_survey2025,
  author  = {M. Sarwar and M. Rizwan and M. Aziz and A. Rehman Sudais},
  title   = {Large language models for power system applications: A comprehensive literature survey},
  journal = {arXiv preprint arXiv:2512.13004},
  year    = {2025}
}

\end{document}